\documentclass[aps,prb,reprint,twocolumn,floatfix,superscriptaddress,amsmath,amssymb,amsfonts,graphicx,nofootinbib]{revtex4-2}

\usepackage{url}
\usepackage{bm}
\usepackage{mathrsfs,dsfont}
\usepackage{amstext}
\usepackage{amsbsy}
\usepackage{verbatim}
\usepackage{color}
\usepackage{floatrow}
\usepackage{float}
\usepackage{tikz}
\usepackage{gensymb}
\usepackage{textcomp}
\usepackage{enumitem}
\usepackage[version=3]{mhchem}
\usepackage{afterpage}
\usepackage{pbox}
\usepackage{siunitx}
\usepackage{stackengine,wasysym,scalerel}
\usepackage{latexsym}
\usepackage{cancel}
\usepackage{comment}
\usepackage{array, multirow, bigdelim, makecell, booktabs}
\usepackage[misc]{ifsym}
\usepackage{times}

\usepackage[normalem]{ulem}

\definecolor{darkblue}{HTML}{004D6B}
\definecolor{darkred}{HTML}{8c1515}
\definecolor{darkgreen}{HTML}{006400}
\usepackage{hyperref}
\hypersetup{
	colorlinks=true,
	urlcolor=darkblue,
	citecolor=darkblue,
	linkcolor=darkblue,
	breaklinks
}
\usepackage[capitalise]{cleveref}

\def\be{\begin{equation}}
\def\ee{\end{equation}}
\def\bea{\begin{eqnarray}}
\def\eea{\end{eqnarray}}

\newcommand{\expectval}[1]{\langle #1\rangle}

\begin{document}

\title{Quantum phase diagram of the spin-$\frac{1}{2}$ Heisenberg antiferromagnet on the square-kagome lattice:\\ a tensor network study}

\author{Saeed S. Jahromi}
\email{saeed.jahromi@iasbs.ac.ir}
\affiliation{Department of Physics, Institute for Advanced Studies in Basic Sciences (IASBS), Zanjan 45137-66731, Iran}
\affiliation{Department of Physics and Quantum Centre of Excellence for Diamond and Emergent Materials (QuCenDiEM), Indian Institute of Technology Madras, Chennai 600036, India}

\author{Yasir Iqbal}
\email{yiqbal@physics.iitm.ac.in}
\affiliation{Department of Physics and Quantum Centre of Excellence for Diamond and Emergent Materials (QuCenDiEM), Indian Institute of Technology Madras, Chennai 600036, India}

\begin{abstract}
We study the ground-state phase diagram of the spin-$1/2$ antiferromagnetic Heisenberg model on the square-kagome lattice using infinite projected entangled-pair states (iPEPS). By systematically varying the ratio of exchange couplings on triangular and square plaquettes, we establish a complete quantum phase diagram in the thermodynamic limit. In the intermediate-coupling regime, we identify four distinct nonmagnetic phases that are unambiguously characterized as valence-bond crystals (VBCs) by their symmetry-inequivalent patterns of strong and weak spin-spin correlations. These include a plaquette crossed-dimer VBC, a loop-six VBC stabilized around the isotropic point, a generalized pinwheel VBC with reduced rotational symmetry, and a decorated loop-six VBC proximate to ferrimagnetic order. We determine the phase boundaries using a combination of bond-resolved correlation functions, entanglement entropy, and magnetization. For transitions not accompanied by sharp entanglement signatures, we extract the spin gap from finite-field simulations, allowing us to distinguish gapped and gapless VBC phases. At larger coupling ratios, the system undergoes transitions into imperfect and perfect ferrimagnetic states. Our results resolve long-standing ambiguities in the square-kagome Heisenberg model and provide a quantitatively reliable reference for future theoretical and experimental studies of frustrated quantum magnets.
\end{abstract}

\maketitle
 
{\it Introduction}. Geometric frustration underpins the emergence of novel quantum phases of matter, which includes a large class of quantum paramagnets such as quantum spin liquids~\cite{Savary2017,Balents2010} described within the resonating valence bond paradigm \cite{Anderson-1973,Moessner2001} and valence bond crystals (VBCs) \cite{Balents2010}. Of particular interest are lattices composed of a corner-sharing arrangement of triangles, since here, antiferromagnetic Heisenberg interactions between spins conspire with geometric frustration to induce a non-trivial infinite degeneracy of classical ground states~\cite{Chalker-1992,Huse-1992,Reimers-1993,Richter-2009}. In such scenarios, quantum fluctuations at $T=0$ often fail to lift this degeneracy giving rise to strongly correlated nonmagnetic phases. Indeed, on the celebrated kagome lattice, the spin $S=1/2$ Heisenberg antiferromagnet represents a classic problem~\cite{Marston1991a,mila_low-energy_1998,Hastings2000,Nikolic2003,singh2007ground,Yan2011,Depenbrock2012}, having been the focal point of rigorous theoretical and experimental investigations~\cite{Khuntia-2020} aimed at characterizing the true nature of its ground state, which according to the latest state-of-the-art numerical approaches, is believed to be a U(1) Dirac spin liquid~\cite{Ran2007,Iqbal2013,Iqbal-2014,He-2017,Liao2017}. 

\begin{figure}
\centerline{\includegraphics[width=\columnwidth]{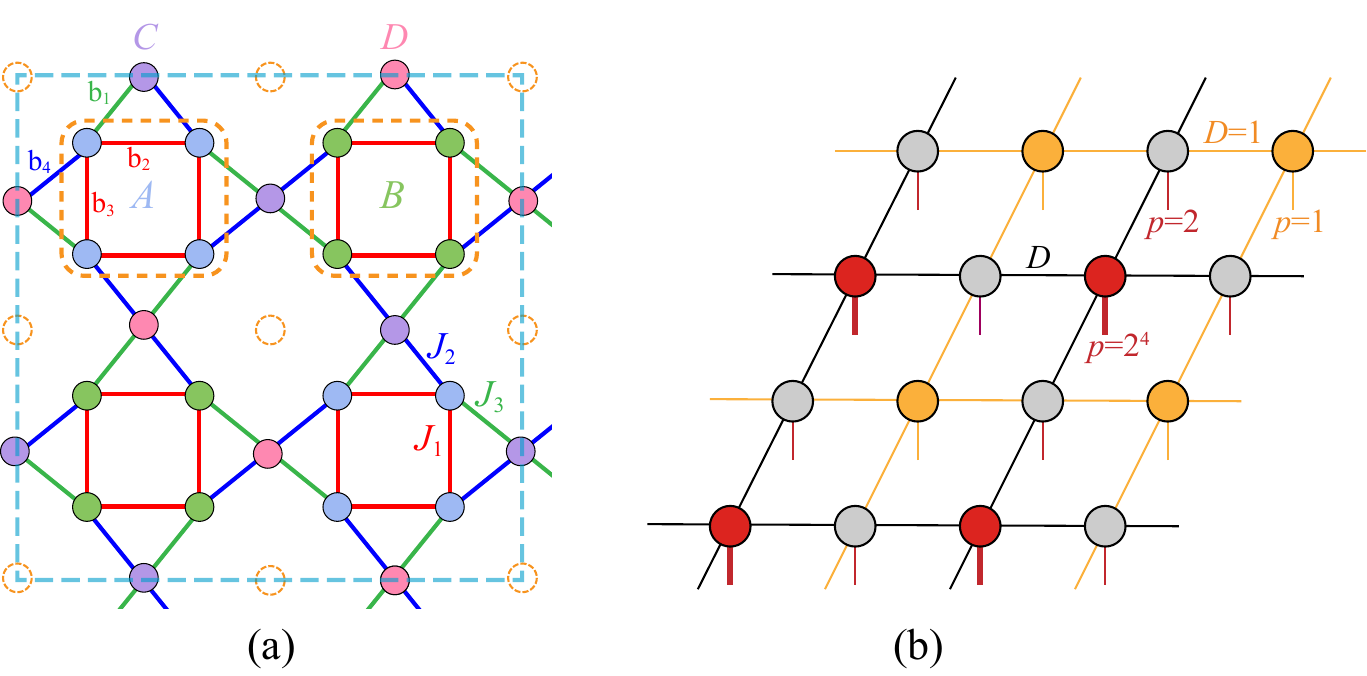}}
\caption{(a) The translation invariant unit cell of the square-kagome lattice with $N_s=24$ physical sites. The blue and green sites represent the two effective sublattices, $A$ and $B$, formed by squares. The purple and pink vertices further represent another set of sublattices, $C$ and $D$, formed by triangles. Additionally, the links ($b_1$, $\ldots$, $b_4$) correspond to the bonds for which the spin-spin correlations have been shown in Fig.~\ref{Fig:phase_details}(b). (b) The corresponding coarse-grained iPEPS tensor networks with square geometry. The dummy virtual indices are colored in orange with $D=1$. The iPEPS tensors are not optimized along these dummy links. The black links further characterize the tensor legs that carry entanglement degrees of freedom.}
\label{Fig:lattice}
\end{figure}

The quest for theoretical and experimental realization of QSLs has recently garnered considerable attention towards quantum magnets exhibiting a novel kagome-like structure referred to as the shuriken or square-kagome lattice~\cite{Siddharthan-2001} [see Fig.~\ref{Fig:lattice}]. It is a four coordinated lattice formed by corner-sharing triangles with two symmetry-inquivalent nearest-neighbor bonds, and two sets of symmetry-inequivalent sites which makes it a non-Archimedean lattice. The field of quantum materials has recently expanded with the identification of new compounds such as KCu$_6$AlBiO$_4$(SO$_4$)$_5$Cl~\cite{fujihala_gapless_2020,Goto-2024}, KCu$_7$TeO$_4$(SO$_4$)$_5$Cl~\cite{Markina2024}, Na$_6$Cu$_7$BiO$_4$(PO$_4$)$_4$[Cl,(OH)]$_3$ \cite{Yakubovich2021,liu_low-temperature_2022} and ACu$_7$TeO$_4$(SO$_4$)$_5$Cl (A=Na, K, Rb, Cs)~\cite{Murtazoev-2023,Rebrov-2024}, where the $S=1/2$ Cu$^{2+}$ ions form essentially decoupled layers of square-kagome lattices. Most interestingly, recent experimental observations suggest that some of these compounds exhibit no sign of magnetic ordering down to $50$ mK with indications for the formation of a gapless quantum spin liquid in KCu$_6$AlBiO$_4$(SO$_4$)$_5$Cl~\cite{fujihala_gapless_2020}, quantum paramagnet in Na$_6$Cu$_7$BiO$_4$(PO$_4$)$_4$[Cl,(OH)]$_3$~\cite{Yakubovich2021,liu_low-temperature_2022,niggemann_requiremhchemquantum_2023} and RbCu$_7$TeO$_4$(SO$_4$)$_5$Cl~\cite{Murtazoev-2023}, and a field induced quantum spin liquid in KCu$_7$TeO$_4$(SO$_4$)$_5$Cl~\cite{Markina2024,gonzalez-2024}. Furthermore, theoretical investigations of the $S=1/2$ Heisenberg antiferromagnetic model, Eq.~\eqref{eq:H_AFH}, with bond anisotropic exchange interactions ($J_1$, $J_2$, $J_3$) [see Fig.~\ref{Fig:lattice}(a)] on the square-kagome lattice suggest that the ground state phase diagram has the potential to accommodate a variety of nonmagnetic phases~\cite{morita_magnetic_2018}, including topological spin liquids (possibly nematic)~\cite{lugan_topological_2019}, VBCs~\cite{rousochatzakis_frustrated_2013,Ralko-2015,schmoll_tensor_2023}, and commensurate as well as incommensurate orders~\cite{lugan_topological_2019}. However, given the limited scope of theoretical approaches employed thus far, there is no consensus regarding the nature and extent of the phases and associate phase transitions, especially in the intermediate coupling regime which is theoretically most challenging, and the quantum phase diagram remains {\it terra incognita}.

Thus, motivated by the theoretical scope of the model in realization of exotic nonmagnetic phases and its possible experimental relevance, we investigate the $S=1/2$ Heisenberg AFM model on the square-kagome lattice along the isotropic line $J_2=J_3\equiv J$. The $J_{1}$\textendash$J$ phase diagram is mapped using large-scale tensor-network (TN)~\cite{Orus2014,Orus2014a,Verstraete2008} simulations based on infinite projected entangled-pair states (iPEPS) {\it Ansatz}~\cite{Orus2009,Phien2015,jahromi_infinite_2018} which allows us to accurately incorporate the subtle interplay between long-range entanglement and short-distance energetics which has been argued to be crucial in reliably ascertaining the precise nature of the quantum phases~\cite{schmoll_tensor_2023}. This is achieved via an accurate estimation of the ground state energy together with a careful analysis of the magnetization and spin-spin correlations, as well as entanglement entropy. 
In addition to the previously reported ferrimagnetic state that emerges in the $J/J_{1} \gg 1$ regime, our study uncovers a rich variety of VBC orders within the intermediate-coupling regime. These VBC phases are unambiguously identified through the distinct symmetry patterns of strong and weak bond correlations obtained from our large-scale iPEPS simulations. To further characterize their nature, we compute the spin gap at representative points of the phase diagram by introducing a Zeeman field and monitoring the onset of magnetization, thereby distinguishing their gapped or gapless nature. Finally, we complement this analysis with a detailed discussion of the symmetry properties of the VBC states and the spontaneous symmetry breaking mechanisms that differentiate them, providing a coherent and quantitative picture of the phases and the transitions connecting them.

{\it Model and Method}. The nearest-neighbor antiferromagnetic Heisenberg model on the square-kagome lattice is
\be
\label{eq:H_AFH}
\mathcal{\hat H} = J_{1}\sum_{\langle i j \rangle_{1}} \mathbf{\hat S}_i \cdot \mathbf{\hat S}_j + J\sum_{\langle i j \rangle_{2}} \mathbf{\hat S}_i \cdot \mathbf{\hat S}_j\,, 
\ee
where, $\mathbf{\hat S}_i$ are $S=1/2$ operators, and the sums $\langle ij\rangle_{1}$ and $\langle ij\rangle_{2}$ run over the two symmetry inequivalent nearest-neighbor links of the lattice, namely, the square and triangular bonds, respectively. While the generic model involves three distinct couplings as illustrated in Fig.~\ref{Fig:lattice}(a), here, we fix $J_{2}=J_{3}\equiv J$. We set $J_{1}=1$ as the energy scale and explore the phase diagram as a function of $J$ within the range $0\leq J\leq 2$.

In order to simulate the $S=1/2$ antiferromagnetic Heisenberg (AFH) model on the square-kagome lattice, we employ the infinite version of the iPEPS algorithm \cite{Orus2009,Phien2015,jahromi_infinite_2018}, specifically adapted for infinite square lattices. In addition, to obtain the tensor network representation of the wave function, we utilized the simple-update algorithm \cite{Jiang2008} equipped with second-order Suzuki-Trotter decomposition for the evolution operators. To approximate the complete contraction of the 2D iPEPS tensor network for the computation of variational energies and expectation values, the corner transfer-matrix renormalization group (CTMRG) method is used~\cite{Nishino1996,Orus2009}. 

To adapt the iPEPS algorithm designed for square lattices directly to the square-kagome structure and approximate its ground state wave function with a square iPEPS tensor network, we initially coarse-grain the square-kagome lattice \cite{jahromi_spin-inf2inf1_2020, schmoll_tensor_2023}. This involves associating a local PEPS tensor with a physical dimension of $p=2^4$, representing the four spins within a square plaquette of the square-kagome lattice [depicted as red tensors in Fig.~\ref{Fig:lattice}(b)]. Subsequently, we assign a distinct PEPS tensor with a physical dimension of $p=2$ to the remaining sites located at the vertices of the triangles [depicted as gray tensors in Fig.~\ref{Fig:lattice}(b)]. Finally, to achieve a fully translation invariant iPEPS network with a square geometry, we introduce dummy tensors with trivial physical and virtual indices at the center of each octagon [represented by yellow tensors in Fig.~\ref{Fig:lattice}(b)]. The original square-kagome lattice and its corresponding TN for the highlighted unit cell are illustrated in Fig.~\ref{Fig:lattice}. Although the smallest allowed TN unit cell for achieving a fully translation invariant square-kagome lattice is $2\times 2$, we conducted our TN simulations using a $4\times 4$ unit cell (equivalent to $N_s=24$ physical sites on the original square-kagome lattice). This choice was made to ensure detection of all potential phases characterized by nontrivial patterns of spin-spin correlations within the unit cell. 

The maximum dimension of virtual bonds achievable within our available computational resources is $D_{\text{max}}= 12$. Furthermore, we set the CTMRG boundary dimension to $\chi=D^2$. For bond dimensions $D\geq 8$, we fix $\chi=64$ due to resource constraints. Nevertheless, we observed that this choice sufficed to obtain converged expectation values for various bond dimensions. Furthermore, we used imaginary time evolution during the simple update, initiating with $\delta\tau=10^{-1}$ and progressively reducing it to $10^{-3}$ while allowing a maximum of $3000$ iterations for each $\delta\tau$. Additionally, we ensured algorithm convergence by monitoring both the energy and singular values obtained during the simple update, terminating the process once a threshold of $10^{-16}$ was reached.  

\begin{figure*}
\centerline{\includegraphics[width=18cm]{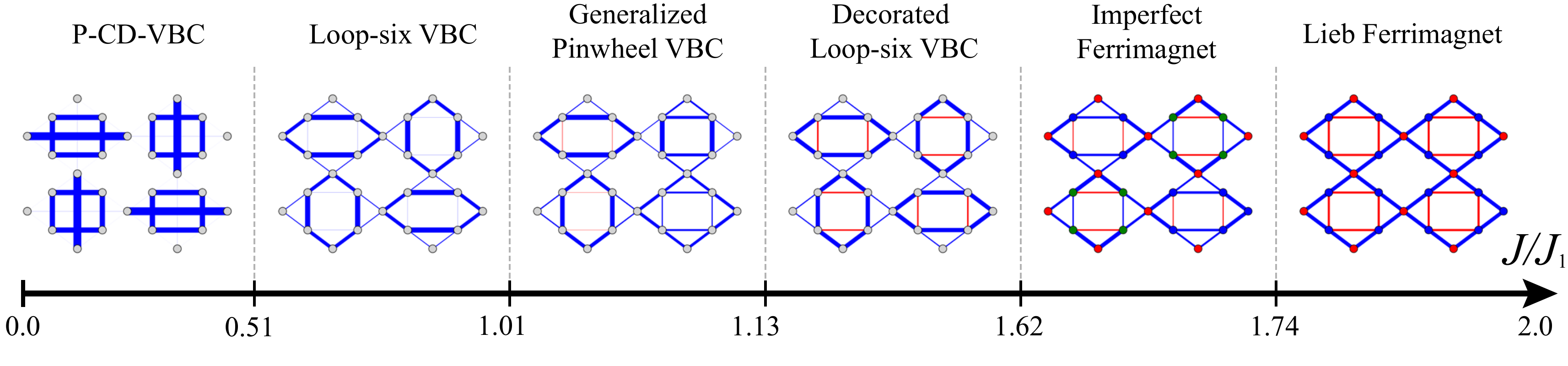}}
\caption{Quantum phase diagram of the $S=1/2$ antiferromagnetic Heisenberg model~\eqref{eq:H_AFH} on the square-kagome lattice. The bond thicknesses are quantitatively proportional to the nearest-neighbor spin–spin correlations $\expectval{\mathbf{\hat S}_i \cdot\mathbf{\hat S}_j}$ extracted from the converged iPEPS wave function at bond dimension $D=12$. Additionally, the blue (red) bonds represent antiferromagnetic (ferromagnetic) correlations with negative (positive) values. In the ferrimagnetic phases, the red sites corresponds to a positive total on-site magnetization value, while the blue, and green sites correspond to negative magnetization, with a different strength.}
\label{Fig:phase_diag}
\end{figure*}

\begin{figure*}
\centerline{\includegraphics[width=\columnwidth]{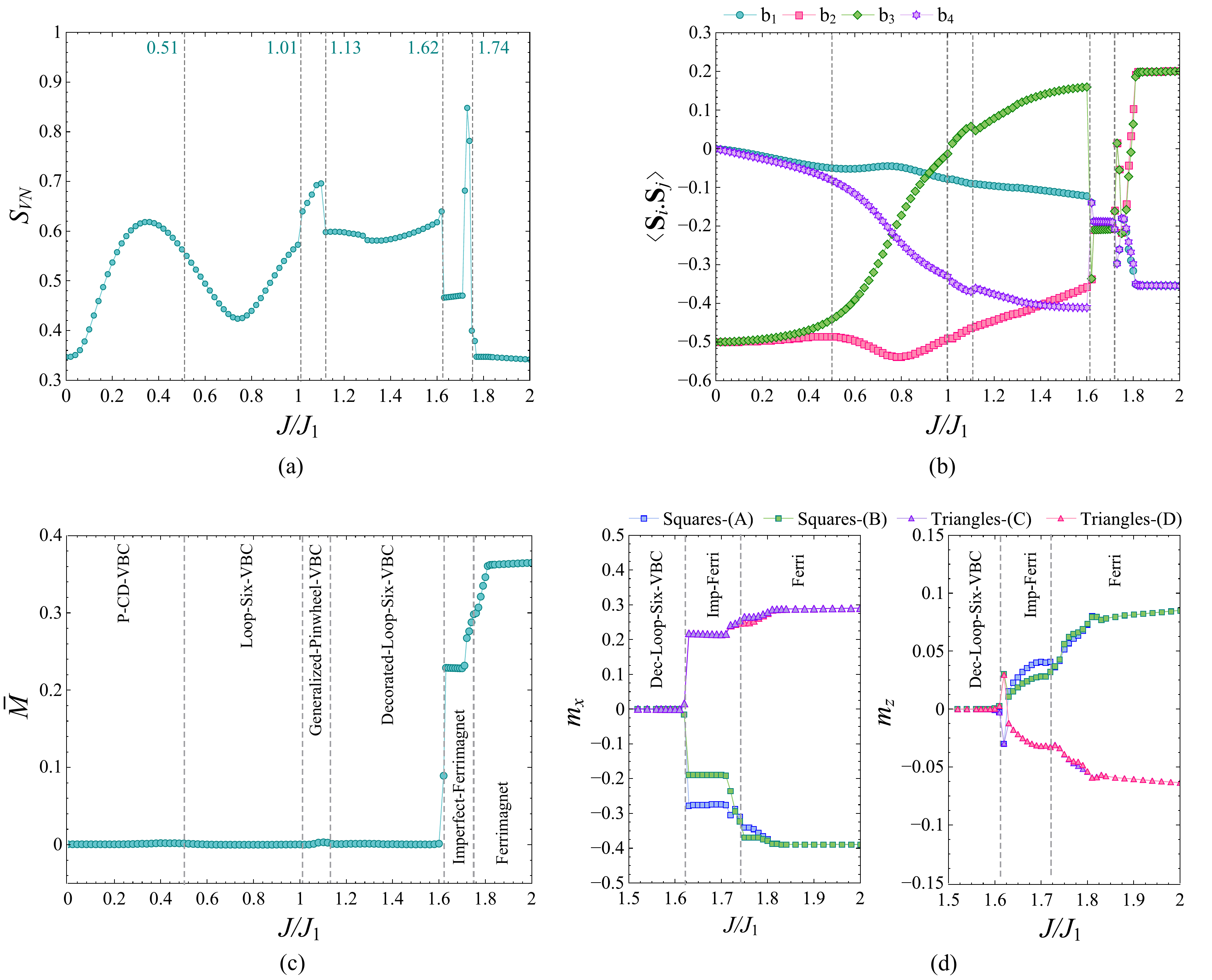}}
\caption{(a) The von Neumann entanglement entropy $S_{VN}$, computed from the singular-value bond matrices of the iPEPS simple-update scheme. Vertical dashed lines mark the phase boundaries  separating the distinct phases in the quantum phase diagram. All phase boundaries are reliably identified by abrupt changes in $S_{VN}$ at the corresponding dashed lines, except for the transition at $0.51$, which is instead identified through an analysis of the spin gap shown in Fig.~\ref{Fig:gap_opening}. (b) The equal-time isotropic spin-spin correlations $\langle\mathbf{\hat S}_i \cdot \mathbf{\hat S}_j\rangle$, for the four bonds [see Fig.~\ref{Fig:lattice}(a) for the link labels] connected to the top-left sites of square $A$ in Fig.~\ref{Fig:lattice}(a). (c) Average total magnetization, $\bar{M} = (1/N_s) \sum_i M_i$, with $M_i = \sqrt{(m_i^x)^2 + (m_i^y)^2 + (m_i^z)^2}$ calculated across all lattice sites. (d) A zoom-in of the average magnetization components, $m_x$ and $m_z$, on the square and triangular sublattices of the unit cell [see Fig.~\ref{Fig:lattice}(a) for the labeling convention]. All the results in the plot correspond to $D=12$ bond dimension.}
\label{Fig:phase_details}
\end{figure*}

{\it Quantum Phase Diagram}.
Using our tailored iPEPS algorithm, we simulated the ground state of the model~\eqref{eq:H_AFH} in the parameter range $0\leq J/J_1\leq 2.0$, for different bond dimensions $D$. In order to characterize the phases and phase transitions of the model, we calculate the variational expectation values of different operators ranging from energy to local magnetization and spin-spin correlations. Based on a careful analysis of the evolution of these quantities together with the Von Neumann entanglement entropy, we identify {\it six} distinct phases in the thermodynamic limit. The quantum phase diagram is shown in Fig.~\ref{Fig:phase_diag} and the iPEPS results supporting the characterization of the underlying phases and phase boundaries are presented in Fig.~\ref{Fig:phase_details} for the largest bond dimension $D=12$.

The phase boundaries between different nonmagnetic phases are most accurately detected from an assessment of the Von Neumann entanglement entropy, $S_{VN} = \sum_i \lambda_i^2 \log\lambda_i^2$, where $\lambda_i$'s are the singular values along the bonds which are obtained from the simple update of PEPS tensors [see Fig.~\ref{Fig:phase_details}(a)]. In addition, when the entanglement-entropy curve varies smoothly, the phase boundaries are further inferred from an analysis of the spin gap. On the other hand, the phase transitions out of the nonmagnetic phases are distinguished from abrupt changes in the average spin-spin correlations [see Fig.~\ref{Fig:phase_details}(b)] as well as the total magnetization and its individual components [see Figs.~\ref{Fig:phase_details}(c) and (d)]. As shown in Fig.~\ref{Fig:phase_details}, in total we observe five transition points separating six phases with distinct patterns of symmetry breaking.  

\begin{figure}
\centerline{\includegraphics[width=\columnwidth]{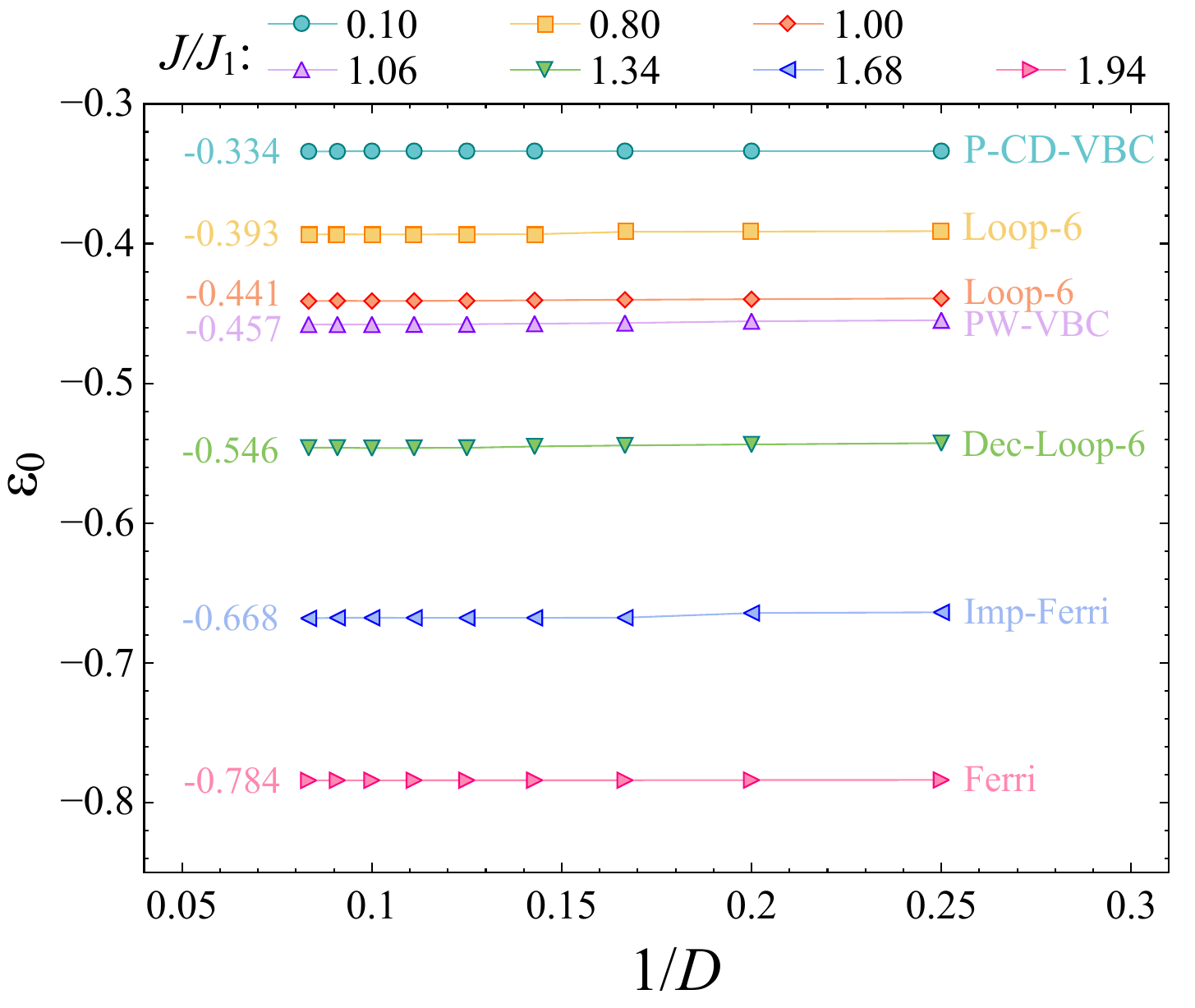}}
\caption{Scaling of the energy per site versus the inverse bond dimension $1/D$ for selected points within each phase. The evident convergence of TN simulations with respect to bond dimension is observed across all phases. The numerical values on the left side of each curve correspond to ground state energy per site, $\varepsilon_0$, for $D = 12$ at the selected point within each phase.}
\label{Fig:e0_scaling}
\end{figure}

The competition between the $J$ and $J_1$ couplings gives rise to broadly two distinct regions in the phase diagram: (i) $0\leq J/J_1\leq 1.62$: which is nonmagnetic composed of an ensemble of VBCs with different symmetries (or by the distinct patterns of which bonds condense) emerging as the ground state of the system and (ii) $J/J_1 > 1.62$: which hosts a ferrimagnetically ordered ground state. In order to characterize the nature of these phases, we compute the local magnetization and spin-spin correlations on individual sites and bonds of the lattice, respectively. Within each region, the pattern of strong and weak bonds in real-space as inferred via equal-time isotropic spin-spin correlations in the ground states is illustrated in Fig.~\ref{Fig:phase_diag}. Each phase is uniquely characterized by a distinct pattern of bond correlations or magnetic ordering, as measured on the converged ground state wave function. It is crucial to emphasize that the convergence of the wave function has been thoroughly examined through various measures within the algorithm. As exemplified in Fig.~\ref{Fig:e0_scaling}, the ground state energy per-site, $\varepsilon_0$, for selected points within each region of the phase diagram exhibits excellent convergence with respect to the inverse bond dimension. We have further characterized the gapped or gapless nature of each phase by computing the spin gap. This was achieved by applying a small Zeeman field $h_z$, tracking the field dependence of the net magnetization $\bar{M}_z(h)$, and determining the critical field $h_c$ at which $\bar{M}_z(h)$ becomes finite. The value of $h_c$ then provides a direct and quantitative measure of the spin gap. The details of spin gap calculations have been provided in the supplementary materials \cite{supp}.

To quantify valence-bond crystalline (VBC) order in the nonmagnetic regime, we emphasize that the relevant order parameters are the spatial patterns of nearest-neighbor bond energies, or equivalently the equal-time spin--spin correlations $\langle \mathbf{S}_i \cdot \mathbf{S}_j \rangle$. These bond observables transform nontrivially under the lattice space-group operations and therefore provide a direct diagnosis of spontaneous breaking of translation, rotation, and/or reflection symmetries. Each VBC phase identified in Fig.~\ref{Fig:phase_diag} corresponds to a symmetry-inequivalent set of bond expectation values within the chosen unit cell, yielding a complete local characterization of the ordered state. In this way, Lieb--Schultz--Mattis--type constraints are resolved through lattice-symmetry breaking, without requiring a symmetry-preserving topologically ordered ground state. We note that nonlocal string or brane order parameters, as employed in one-dimensional spin chains or in two-dimensional Mott insulating and gauge-theoretic settings, are designed to diagnose symmetry-preserving or topological phases and are therefore not the natural diagnostics for valence-bond crystalline states characterized by explicit lattice-symmetry breaking in local bond observables~\cite{Chen-2008,Rath-2013,Fazzini-2017,Fazzini-2019}.

\begin{figure}
\centerline{\includegraphics[width=\columnwidth]{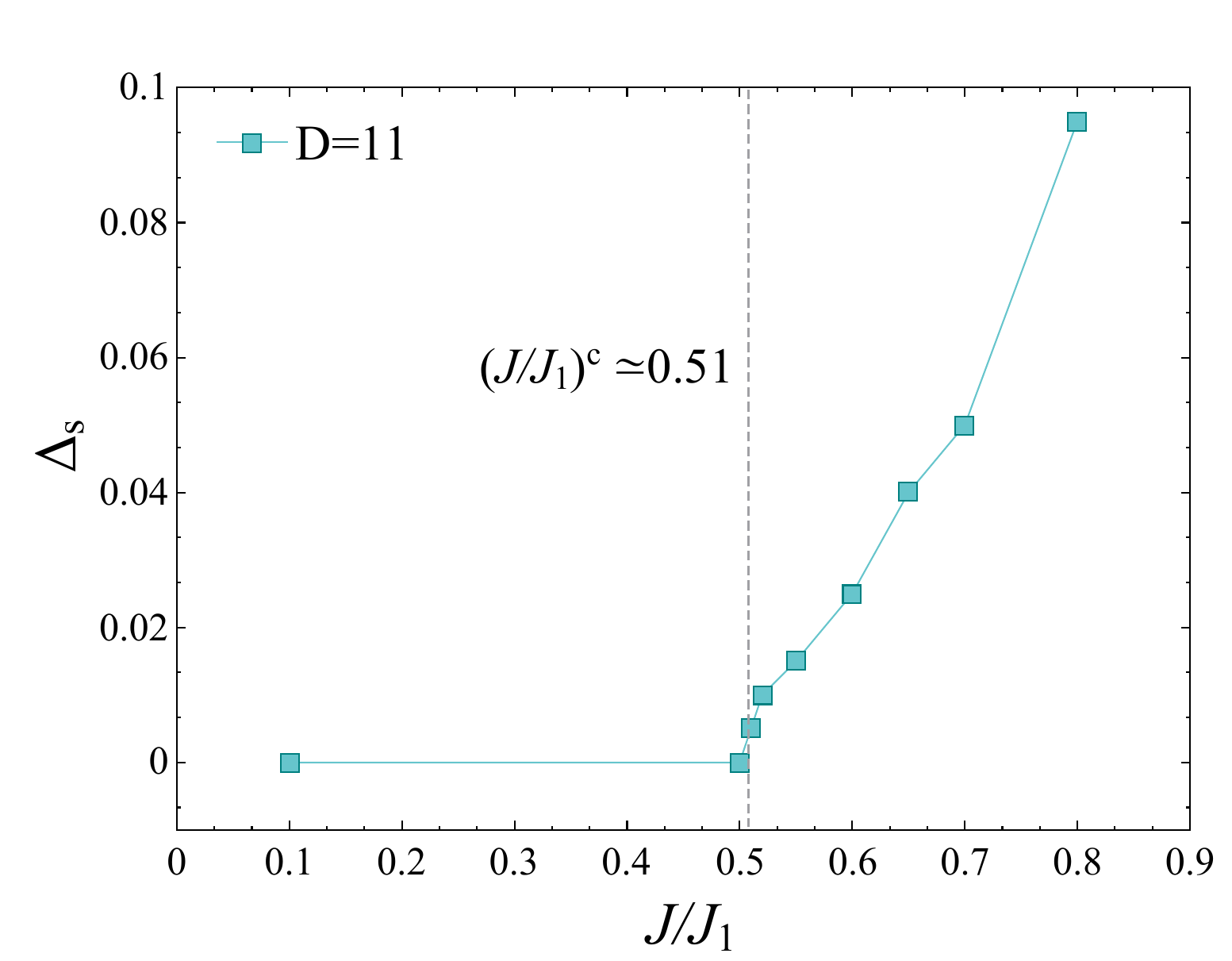}}
\caption{The spin gap across the transition from the P-CD-VBC to the loop-six VBC phase. A finite gap opens at $J/J_1\approx 0.51$, indicating a transition from a gapless P-CD-VBC to a gapped loop-six VBC phase. The data are obtained from iPEPS simulations with bond dimension $D=11$.}
\label{Fig:gap_opening}
\end{figure}

We now elaborate on the nature of each of these phases. At $J/J_1=0$, we have isolated squares with antiferromagnetically interacting spins. Thus, the ground state of the system is plaquette ordered with the four spins forming isolated tetramers with total spin zero. In the limit $J/J_1\to0$, we find that the spins at the triangular apex sites are not free but rather form long-range dimers in a checkerboard arrangement [see Fig.~\ref{Fig:phase_diag}]. This phase, with long-range singlet formation triggered by a nonzero $J$ coupling, was predicted within a perturbative treatment of the model~\cite{rousochatzakis_frustrated_2013}, and referred to as plaquette crossed-dimer VBC (P-CD-VBC). Here, the bond correlators $\langle \mathbf S_i\!\cdot\!\mathbf S_j\rangle$ (equivalently bond energies) are $C_4$ symmetric. The P-CD-VBC state is found to be stable till $J/J_1 \lesssim 0.51$ when it transitions to a loop-six VBC~\cite{Ralko-2015} which has the same $C_4$ symmetry, and is thus distinguished by which bonds condense. The transition point is thus identified through an analysis of the spin gap shown in Fig.~\ref{Fig:gap_opening}. One can clearly see that a finite spin gap opens for $J/J_1 \gtrsim 0.51$, signaling a continuous phase transition from a gapless phase to a gapped state. Moreover, at the transition point, the spin–spin correlations on bonds $b_1$, $\ldots$, $b_4$ begin to deviate from one another, forming a characteristic loop-six pattern [see Fig.~\ref{Fig:phase_details}(b)].
In the loop-six VBC~\cite{Ralko-2015}, which is stabilized due to strong resonances over longer-length loops amplified by the dressing of virtual singlets on top of the nearest-neighbor basis, we find that the bonds with strong antiferromagnetic correlation form closed hexagonal loops in an alternating horizontal and vertical checkerboard pattern which is $C_4$ symmetric. The loop-six VBC is stable over a broad parameter range $0.51\leq J/J_1\leq 1.01$ encompassing the isotropic point $J/J_1=1$, where previous studies based on quantum dimer models~\cite{Ralko-2015} and tensor networks~\cite{schmoll_tensor_2023} have presented evidence for its ground state candidature. Interestingly, at $J/J_{1}\sim 1$, in the ED spectra, one notices a change in symmetries of the low excited states, and indeed for $J/J_1 \geq 1.01$, our iPEPS calculations find the appearance of a generalized version of the conventional pinwheel VBC~\cite{rousochatzakis_frustrated_2013} which was analyzed as a competing ground state at the isotropic point in earlier VMC~\cite{astrakhantsev_pinwheel_2021} and tensor network studies~\cite{schmoll_tensor_2023}. This generalized pinwheel state has broken reflection symmetries akin to the conventional pinwheel VBC but lacks the $C_4$ rotation symmetry which gets lowered to $C_2$ [see Fig.~\ref{Fig:phase_diag}], and the consequent emergence of a preferred lattice direction. This symmetry breaking and the corresponding increase in the number of variational parameters is responsible for its energetic stabilization till $J/J_1\leq 1.13$. Finally, the last phase of the VBC family in the phase diagram is a decorated version of the loop-six VBC and emerges within the range $1.13\leq J/J_1\leq 1.62$. This state differs from the conventional loop-six VBC state by the appearance of additional parallel alternating horizontal and vertical ferromagnetic correlations on the edges of the squares, which not only lower its symmetry from $C_4$ to $C_2$, but are signatures of the incipient ferrimagnetic order that is proximate in the phase diagram, and to which this phase transitions into at $J/J_1\sim 1.62$. Again, this interval of the decorated loop-six VBC approximately coincides with a clearly identifiable distinct region in the ED spectra. In summary, our results show that the quantum phase diagram hosts four types of VBC states in the region $0\leq J/J_1\leq 1.62$, which are distinguished by distinctive patterns of strong and weak spin-spin correlations of the links of the squares and triangles of the square-kagome lattice.

Next, by increasing $J/J_1$ beyond $1.62$, we enter the second regime of the phase diagram that is spanned by magnetically ordered ground states with a ferrimagnetic spin arrangements on the square and triangular sublattices. It is around $J/J_{1}\sim1.6$ that one observes a level crossing of states with two different symmetries, which could potentially be linked to this phase transition. In the first phase, which occupies $1.62\leq J/J_1\leq 1.74$, the spins on the squares are anti-aligned with those of triangles. However, different square sublattices, $A$ and $B$, exhibit different magnitudes for both the $m_x$ and $m_z$ components of magnetization [see Fig.~\ref{Fig:phase_details}(d)]. We call this phase an imperfect ferrimagnet due to difference in magnitude of the local magnetization of the square sublattices. However, by increasing the strength of the exchange couplings on triangles we approach the classical limit and transition to a perfect ferrimagnetic state for $J/J_1\geq 1.74$, wherein all spins on squares point in the same direction and with the same magnitude, while they are antiparallel to spins on triangles. This sequence of phases is consistent with a previous exact-diagonalization (ED) study~\cite{morita_magnetic_2018,Richter-2023}. It is important to note that the shift of the transition to a ferrimagnetic phase to smaller values compared to its classical value of $J/J_{1}=2$ is due to the order-by-disorder mechanism which 
stabilizes the collinear up-up-down state at smaller values of $J/J_{1}$. 

This phase is an example of the so-called Lieb ferrimagnet~\cite{lieb_two_1989} on the square-kagome lattice and corresponds to an up–up–down spin configuration, in which the spins on the square and triangular sites are antiparallel. Our iPEPS calculations show that the total spin of the ferrimagnetic ground state is $S_{\text{tot}} = 3.992$, in excellent agreement with the Lieb-theorem prediction $S_{\text{tot}} = \frac{|N_T - N_S|}{2}=4$, where $N_T = 8$ and $N_S = 16$ denote the number of sites on the triangular and square sublattices of the $24$-site unit cell shown in Fig.~\ref{Fig:lattice}(a). Moreover, our numerical results demonstrate that the total spin satisfies the condition $0<S_{\text{tot}}<\frac{1}{3}M_{\text{sat}}$, where $M_{\text{sat}} = \frac{N_T + N_S}{2}=12$ is the saturated magnetization of the $24$-site unit cell, resulting in a finite net magnetization with $S_{\text{tot}}/M_{\text{sat}} = 1/3$. The observed ferrimagnetic state underscores the consistency of our simulations with the Lieb ferrimagnetic framework and further validates the robustness of our TN simulations for capturing the intricate quantum and classical properties of the ferrimagnetic phases.

Lastly, we investigated the gapped or gapless nature of the phases by computing the spin gap through the application of a small Zeeman field $h_z$ and tracking the field dependence of the net magnetization $\bar{M}_z(h)= \sum_i \expectval{S_i^z}/(N_s S)$ for several representative points in the phase diagram [see supplementary materials \cite{supp}]. The critical field $h_c$ at which $\bar{M}_z(h)$ becomes finite provides a direct measure of the spin gap. Our results reveal that the P-CD-VBC phase and the ferrimagnetic phases are gapless in the spin-triplet sector (i.e., vanishing spin gap), as indicated by their immediate response to an infinitesimal field ($h_c=0$). In contrast, the loop-six, generalized pinwheel, and decorated loop-six VBC phases exhibit finite spin gaps with non-vanishing $h_c$ values. In particular, the loop-six VBC shows $h_c=0.095$ at $J/J_1=0.8$ which gradually decreases to $h_c=0.041$ at the isotropic point $J/J_1=1.0$ consistent with previous TN studies \cite{schmoll_tensor_2023}. Upon entering the generalized pinwheel phase, the gap narrows further to $h_c=0.017$ at $J/J_1=1.06$ before reopening substantially in the decorated loop-six VBC with $h_c=0.112$ at $J/J_1=1.34$, indicating a robust gapped state proximate to the ferrimagnetic phase. These results provide a clear energetic distinction between gapped and gapless phases and complement the real-space characterization of the phase diagram.

Because neighboring phases have different residual space-group symmetries, their boundaries correspond to symmetry-changing transitions. In our data, these are accompanied by abrupt rearrangements of the bond pattern and features in $S_{\mathrm{VN}}$, consistent with weakly first-order or continuous transitions depending on the specific phase boundary.
 
{\it Conclusions}. Using large-scale tensor network calculations based on the infinite projected entangled-pair state algorithm, we present the ground state phase diagram of the $S=1/2$ Heisenberg model on the square-kagome lattice with two symmetry-inequivalent nearest-neighbor antiferromagnetic couplings. Within the accuracy of our approach, we find that the model has a nonmagnetic singlet ground state for $0\leq J/J_{1}\leq1.62$ and a ferrimagnetic ground state for $J/J_{1}\geq1.62$, in agreement with the findings in Ref.~\cite{Richter-2023}. The nonmagnetic region hosts four distinct VBC orders with different patterns of spin-spin correlations. These VBCs can also be viewed as instabilities of parent gapless spin liquids~\cite{Song-2019}, such as those classified for the square-kagome lattice in Ref.~\cite{astrakhantsev_pinwheel_2021}. It would be important to identify the parent spin liquid and to identify the mechanism of dimerization, i.e., fermion bilinear instability or condensation of singlet monopoles~\cite{Budaraju-2023,Budaraju-2024}. This could be assessed within a variational Monte Carlo approach which has been successful in ascertaining the nature of dimer instabilities of gapless spin liquids on the kagome lattice~\cite{Iqbal-2011,Iqbal-2012,Iqbal-2018}.

As a natural continuation for future studies, it would be interesting to investigate the thermodynamics of this model~\cite{Tomczak-2003,Nakano-2013,Derzhko-2014,Hasegawa-2018,Richter-2022} within a tensor network approach~\cite{Schmoll-2024}. It was argued in Ref.~\cite{rousochatzakis_frustrated_2013} that the inclusion of a magnetic field would induce the elusive spin-nematic phases, whose detection within a TN framework would constitute an important step. In search of QSL behavior, it would be worthwhile to assess the effect of quantum fluctuations on the classical phase diagram of the nearest-neighbor $J_{1}$-$J_{2}$-$J_{3}$ model which has extended classical spin liquid and N\'eel ordered phases~\cite{Gembe-2023,Yogendra-2024}. In similar vein, the classical noncoplanar orders triggered by long-range Heisenberg couplings~\cite{Gembe-2023} could potentially melt under strong quantum fluctuations for $S=1/2$ giving rise to chiral spin liquids, which would be worth investigating within a TN framework.
  
{\it Acknowledgements}. We thank Ioannis Rousochatzakis and Arnaud Ralko for helpful discussions. S.S.J. acknowledges support from Institute for Advanced Studies in Basic Sciences (IASBS). S.S.J. thanks IIT Madras for a Visiting Faculty Fellow position during which this work was completed. The work of Y.I. was performed in part at the Aspen Center for Physics, which is supported by National Science Foundation Grant No.~PHY-2210452 and a grant from the Simons Foundation (1161654, Troyer). This research was supported in part by grant NSF PHY-2309135 to the Kavli Institute for Theoretical Physics (KITP).
Y.I. acknowledges support from the ICTP through the Associates Program, from the Simons Foundation through Grant No.~284558FY19, IIT Madras through the Institute of Eminence (IoE) program for establishing QuCenDiEM (Project No.~SP22231244CPETWOQCDHOC), and the International Centre for Theoretical Sciences (ICTS), Bengaluru during a visit for participating in the program: Kagome off-scale (ICTS/KAGOFF2024/08).
Y.~I.~acknowledges the use of the computing resources at HPCE, IIT Madras. 

\clearpage
%
\clearpage

\newcommand{\addpage}[1] {
\begin{figure*}
  \includegraphics[width=8.5in,page=#1]{SuppMat.pdf}
\end{figure*}
}
\addtolength{\oddsidemargin}{-0.75in}
\addtolength{\evensidemargin}{-0.75in}
\addtolength{\topmargin}{-0.725in}
\addpage{1}
\addpage{2}

\end{document}